\documentclass[reprint,aps,prstper]{revtex4-2}

\usepackage{graphicx}
\usepackage{hyperref}
\usepackage{footnote}
\usepackage{amsmath,amssymb}
\usepackage{multirow}
\usepackage{tabularx}
\usepackage{array}
\usepackage[utf8]{inputenc}
\usepackage{enumitem}
\usepackage{subfig} 

\begin{document} 

\title{AI and the FCI: Can ChatGPT project an understanding of introductory physics?}

\author{Colin G. West$^1$}
\email[]{colin.west@colorado.edu}
\affiliation{$^{1}$Department of Physics, University of Colorado, Boulder, Colorado 80309, USA}

\date{\today}

\begin{abstract}
ChatGPT is a groundbreaking ``chatbot"--an AI interface built on a large language model that was trained on an enormous corpus of human text to emulate human conversation. Beyond its ability to converse in a plausible way, it has attracted attention for its ability to competently answer questions from the bar exam and from MBA coursework, and to provide useful assistance in writing computer code. These apparent abilities have prompted discussion of ChatGPT as both a threat to the integrity of higher education and conversely as a powerful teaching tool. In this work we present a preliminary analysis of how two versions of ChatGPT (ChatGPT3.5 and ChatGPT4) fare in the field of first-semester university physics, using a modified version of the Force Concept Inventory (FCI) to assess whether it can give correct responses to conceptual physics questions about kinematics and Newtonian dynamics. We demonstrate that, by some measures, ChatGPT3.5 can match or exceed the median performance of a university student who has completed one semester of college physics, though its performance is notably uneven and the results are nuanced. By these same measures, we find that ChatGPT4's performance is approaching the point of being indistinguishable from that of an expert physicist when it comes to introductory mechanics topics.  After the completion of our work we became aware of Ref~\cite{Gerd}, which preceded us to publication and which completes an extensive analysis of the abilities of ChatGPT3.5 in a physics class, including a different modified version of the FCI. We view this work as confirming that portion of their results, and extending the analysis to ChatGPT4, which shows rapid and notable improvement in most, but not all respects.

\end{abstract}

\maketitle


\section{Introduction} 
 ``ChatGPT," in simplest terms, is a software application designed to mimic human conversation by producing and responding to text, a skill called ``natural language processing."~\cite{OpenAI} Technically, it is based on a ``large language model" (LLM)  which makes use of two recent advances in the LLM field: the ``Transformer" model~\cite{TransformerOriginal} and ``pretraining."~\cite{Pretraining} from whence arises ``GPT" (it is a [G]enerative, [P]retrained [T]ransformer model). While a great deal has been written about the methods used to produce the ChatGPT system~\cite{FewShot}, it suffices here to note that it is one of a new generation of artificial language processing systems-- sometimes colloquially called ``chatbots"-- which has garnered substantial attention in both academic~\cite{Importance1} and popular press~\cite{Importance2} for its ability to seemingly carry on a coherent conversation and complete other tasks. There are two versions of the model available to researchers at the time of this writing: ChatGPT3.5, which is available to the public, and ChatGPT4, which is available only under paid subscription and with substantial limitations on usage rates. The latter model is intended to deliver significantly superior performance than its predecessor~\cite{GPT4}. In this paper, we will write "ChatGPT" when making statements that apply equally to both versions, and add the version numbers in other cases. 

Recent papers have shown that, by at least some measures, ChatGPT's ability to converse like a human also allows it to seemingly display competence in fields like business and law. For example, its responses to questions about Operations Management, a core topic in many MBA programs, were assessed in one study as being at the "B or B-" level.~\cite{wharton} Another work concluded that, ``although ChatGPT[3.5] would have been a mediocre law student, its performance was sufficient to successfully earn a JD degree from a highly selective law school"~\cite{lawschool}. A similar work projected that, given the surprisingly strong performance of ChatGPT3.5 on sample questions it could not have seen before, a similar LLM might be able to pass an actual bar exam ``within the next 0-18 months."~\cite{barexam}. ChatGPT4 appears to have reached that benchmark approximtely three months later~\cite{BarExam4}. The improved model also scores at very high levels--routinely at the 80th or 90th percentile or above--on many standardized tests, including notably the AP Physics C exam~\cite{GPT4}.

A central premise of much physics education work is that there is a difference between the ability to solve classically formulated ``physics problems" and an ability to demonstrate true conceptual ``understanding." The field of Physics Education Research (PER) has developed tools specifically to probe this distinction, and here, we deploy one of the fields most venerable instruments, the Force Concept Inventory (FCI)~\cite{FCI} to compare its apparent conceptual performance with typical students from an introductory physics course at the college level. 

We note that the work of Ref~\cite{Gerd}, which was made available several weeks before this work, also includes an analysis of ChatGPT3.5 using the FCI. Since our works were completed without knowledge of one another and since ChatGPT's responses are inherently probabilistic, we view this work as serving as a confirmation of some of their results: namely, that ChatGPT can correctly answer roughly 60 to 65\% of the questions on the FCI, and that it's answers are usually but not always relatively stable. These results are discussed in more detail below, including some alternative analysis, and we then extend the work to reconsider the results with ChatGPT4. Certainly the authors of Ref~\cite{Gerd} are likely pursuing similar work which we hope will also serve as useful confirmation or contrast with ours.

\section{Background}

ChatGPT was designed and optimized \textit{specifically} for the art of conversing in a manner that would seem plausibly human~\cite{FewShot}. To a loose approximation, it has analyzed and internalized the patterns of words in an enormous sample of human text (principally the "CommonCrawl" dataset)~\cite{FewShot}), which consists of things like books, news articles, wikipedia pages, reddit threads, and content from more specialized and technical fora like StackExchange and StackOverflow~\cite{FewShot, commoncrawl}. From this data, it has an immense probabilistic model of how words tend to be fit together by a human being in various context. For example, it "knows" that a sentence which begins ``after the fight, William hurt..." is likely to end with something like ``...his hand." But it also recognizes that the sentence ``after the fight \textit{scene}, William Hurt..." might instead end with "... appeared to be injured and had to be helped off set." Crucially, this means that ChatGPT is a tool which manipulates and responds to language, and is not designed or trained to implement any model of an underlying concept~\footnote{There may be an exception for raw mathematics--see Methods section} like business, law, or physics. But it is designed to talk like a person, and people who talk about business, law, and physics generally discuss these topics with some intelligence. Hence, even without any specific training in physics, its familiarity with the way physicists talk about physics may be enough for it to project at least an appearance of understanding.

 The concept of``Understanding" can be deep and difficult to define, particularly in the context of learning a new topic~\cite{6facets}. We will generally leave that longstanding question~\cite{minsky1982} to the likes of AI researchers, cognitive scientists and philosophers, and for this reason we have been very careful with the wording in the title of this paper and elsewhere. We are not about to ask, because we are not equipped to know, whether ChatGPT \textit{understands} introductory physics. We ask whether its behavior \textit{creates the appearance of understanding} to the outside world-- which might either be because it has succeeded in understanding, or merely because it has figured out how to display all the usual indicia of understanding. Hence, in the same manner that some poker player might ``project  an aura of confidence" either as a proud display of internal fortitude or as an act desperate bluffing, we consider here whether ChatGPT ``projects" an understanding of introductory physics to the outside world. 

There is a deep body of work in the literature of physics education about how to assess whether a student is demonstrating ``understanding,"~\cite{6facets, Noah}. To start with the basics, we choose as our primary assessment tool a classic instrument which has been used almost as a gold standard for decades: the Force Concept Inventory (FCI)~\cite{FCI}. This influential and heavily-studied assessment is a set of 30 questions designed to try to isolate  and allow students to demonstrate a \textit{conceptual} understanding of introductory kinematics and dynamics, as they might be covered in the first semester of introductory physics at the high school or university level. Although it has plenty of limitations, it was designed as a tool with an eye to distinguishing true conceptual mastery from the kinds of rote memorization, pattern-matching, and algorithmic calculation\cite{CarlKathy, elby, patternmatching} which students sometimes use in order to pass conventional physics tests without ever truly knowing what they are doing, or why. Its status as a classic test for understanding in introductory physics makes it a logical starting point for our inquiry, though we encourage future work to extend beyond this starting point as well. 

In short, we will offer ChatGPT a modified version of the questions from the FCI, and assess on multiple levels how successfully it can project understanding of these topics in intro physics. Section~\ref{sec:methods} of this paper we describe in more detail how the assessment was modified and administered, and in section ~\ref{sec:results} we analyze ChatGPT's responses through various lenses and compare its performance to a large sample of real human students. Section ~\ref{sec:conclusions} gives a summary of our resulting inferences and offers speculations about a number of open questions and directions for future work.

\section{Methods}{\label{sec:methods}
Because all of our reference points for what it looks like when someone projects an understanding of physics are based on the performance of other human beings, Hence, it was important when administering the FCI that we hew as closely as possible to the conditions as our human students would encounter it. This was not entirely possible, and some modifications were necessary both to the questions themselves as well as to the process of giving them to ChatGPT. Such changes were kept to a minimum. In this section we detail those changes and the resulting procedure for assessing ChatGPT with the FCI.

\subsection{Suitability of the FCI}
The FCI is a natural fit as a first step in assessing the capabilities of ChatGPT for several reasons mentioned above: it is focused on conceptual understanding rather than computation and memorization; it has been widely-studied and validated as an assessment tool, and because it has been given frequently to many students in introductory physics, providing natural benchmarks for comparison. But it is also potentially valuable for another reason. To preserve its integrity as an assessment tool, the providers of the FCI have taken steps to encourage practitioners who use it to keep its contents (and even more importantly, its solutions) from becoming widely available. These efforts have certainly not been flawlessly successful but it remains the case that FCI text and solutions are difficult to find on the internet. Where they do exist, they are typically either password-protected on websites used by physics educators or at least ``paywalled" on websites used by students. The relative scarcity of FCI solutions on the internet is consistent with the findings of prior work showing that access to the internet does not undermine the validity of tools like the FCI as conceptual assessment, even when students are observed to be copying question text for the presumable purpose of searching for its answers~\cite{WilcoxPollock}. And in our case, since we are administering the FCI to a program that can only parse text, we can go even further: a significant portion of the FCI problems and/or solutions that can be found online exist as scanned images and/or had-annotated PDFs, which means that they would not be parsed by the kinds of automated tools that scrape the web for text. 

While the exact details of the text which was used to ``train" ChatGPT are a proprietary secret, it is known that its reading material was largely drawn from the ``Common Crawl" corpus~\cite{commoncrawl}, an open repository of data scraped from text found on the public internet. Common Crawl allows users to query which domains it has indexed; we used this feature to verify that it has not indexed the handful of websites which we are most familiar with which might contain solutions to the FCI, either as a PER tool or as a repository of solved problems for students. Beyond the CommonCrawl corpus, it is believed that most of ChatGPT's training data came from specifically generated human  with human feedback~\cite{RLHF}, which of course are highly unlikely to contain references to the FCI. And finally, in the rare locations that we were able to find FCI solutions online, they were typically stored apart from the questions themselves, meaning that there was no obvious reason an LLM would know to pair particular solution texts with particular problems even if it had access to them. For all these reasons, we believe that the FCI is likely \textit{not} in the training text of ChatGPT and hence that its responses have to represent more than regurgitation of something it ``remembers." In this respect our testing with the FCI is what researchers in AI and machine learning might term ``zero shot task": a challenge in which the model is used to classify (and in this case, respond to) prompting text it has never seen before.

\subsection{Modifying the FCI}

The FCI is a 30-item sequential multiple-choice assessment, with each item containing five choices (four distractors and one unique correct answer or ``key").~\cite{FCI} It's items cover a range of topics from approximately the first third of a semester of college-level introductory physics: kinematics, projectile motion, free-fall, circular motion, and Newton's laws. 
This means that generally, it is well-suited to our task. Its one significant drawback is that 18 of its 30 items contain some kind of reference to a figure. ChatGPT3.5 is designed only to accept text input, and despite some clever attempts to feed it images in some sort of indirect or transformed state~\cite{blogpost}, it does not seem capable currently of extracting any meaningful information from a picture. ChatGPT4 advertises multimodal capability which might make it possible to prompt the model with a figure, but this feature is not publically available as of this writing.

Of the 18 items with figures, we find that 11 of them can be modified by adding text that described what was shown in the figures without fundamentally altering the task at hand. In doing so, we take care to make sure that we did not provide additional clues or context that would make the problem simpler for ChatGPT than it would be for a typical physics student. For example, item seven involves a steel ball on a rope being swung in a circular path and then suddenly cut free. The question asks about the path of the ball after it is released, and the figure supplies several different possible trajectories. One way to describe these trajectories in words would be to to say ``tangential to the circle," ``normal to the circle," etc. But we suspect this modification would substantially alter the difficulty of the problem. Instead, we translate this figure into words with reference to cardinal directions: \begin{quote}
    Consider a moment in the ball’s motion when the ball is moving north. At that moment, the string breaks near the ball. Which of the following paths would the ball most closely follow after the string breaks?

\begin{enumerate}[label=(\alph*)]
    
    \item It will initially travel north, but will quickly begin to curve to the west
\item It will travel north in a straight line
\item It will travel northeast in a straight line
\item It will initially travel east, but will quickly begin to curve north
\item It will travel East in a straight line

\end{enumerate}

\end{quote}

We feel that this wording captures exactly and unambiguously all of the different paths indicated in the original figure, but without providing any additional hints. If anything, it may make the item slightly harder for ChatGPT than the original version.

Thirteen other items from the FCI were modified in a similar way. Six items without figures received minor text modifications that should not have affected the nature of the physics being tested. For example, in clusters of questions where some items referenced ``the previous problem," we removed these references and simply restated the set-up from the prior problem, so that items could be asked about independently if needed. We also rephrased any questions that were left with an open-ended statement for the student to complete, since initial experiments showed that ChatGPT3.5 occasionally appeared ``confused" when it was not explicitly asked a question. Hence, a question like item one, which originally read: 

\begin{quote}
    Two metal balls are the same size but one weighs twice as much as the other. The balls are dropped from the roof of a single story building at the same instant of time. The time it takes the balls to reach the ground below will be:

\begin{enumerate}[label=(\alph*)]
    
    \item About half as long for the heavier ball as for the lighter one
    \item About half as long for the lighter ball as for the heavier one
    \item About the same for both balls
    \item Considerably less for the heavier ball, but not necessarily half as long
    \item Considerably less for the lighter ball, but not necessarily half as long

\end{enumerate}

\end{quote}

Was rephrased to end with a direct question: 

\begin{quote}
    Two metal balls are the same size but one weighs twice as much as the other. The balls are dropped from the roof of a single story building at the same instant of time. Which of the following describes the time it takes the balls to reach the ground below?

\begin{enumerate}[label=(\alph*)]
    
    \item About half as long for the heavier ball as for the lighter one
    \item About half as long for the lighter ball as for the heavier one
    \item About the same for both balls
    \item Considerably less for the heavier ball, but not necessarily half as long
    \item Considerably less for the lighter ball, but not necessarily half as long

\end{enumerate}

\end{quote}

It seems unlikely that such changes impacted either the physics content or the difficulty of the items. Finally, four of the items were left entirely unchanged. 

This meant that we were able to ask ChatGPT 23 of the FCI's 30 items. Although others have shown that it is possible to get a representative sample of a student's performance using only a subset of the FCI questions~\cite{HalfFCI}, it happens that the ``unusable" questions are not uniformly distributed across all question categories. Removing problems 19 and 20, for example, meant removing the only questions on linear kinematics from the instrument. Although this affects our ability to make comparisons with results from the ``full" FCI, we believe this difficulty can be overcome, as we shall discuss in Sec~\ref{sec:results} below. 

\begin{table}[ht]\label{table:changes}
\begin{center}
\begin{tabular}{||c | c||} 
 \hline
Type of change & Items\\ [0.5ex] 
 \hline\hline
None & {1, 4, 29, 30}  \\ 
 \hline
 Minor text & { 2, 3, 13, 25, 26, 27 }  \\
 \hline
 Figure description & {5, 7, 9, 10, 11,  15, 16, 17 18, 22, 23, 24, 28} \\
 \hline \hline
 Unusable & {6, 8, 12, 14, 19, 20, 21} \\ [1ex] 
 \hline
\end{tabular}
\end{center}
\caption{Table of items from the FCI and the ways that they were (or were not) modified for use in this work. Seven items were not used.}
\end{table}

\subsection{Administering the FCI to ChatGPT}
We began interacting with the version of ChatGPT which existed during the month of January, 2023, and used our initial explorations there to develop a the guidelines we used for how to pose questions. Instances of conversations with ChatGPT are completely separate, in the sense that ChatGPT does not ``remember" content from one chat in a separate chat, so variations of a question can be asked in parallel to identify the best practices for posing the questions. On the other hand, \textit{within} a conversation ChatGPT can remember content back to a depth of about 3000 words. But in practice, this is not enough to remember a full administration of the FCI. This is part of the reason that we chose to rephrase each question so that it could stand alone, rather than referencing things from ``the previous question," etc. 

Our questions to ChatGPT were each posed in the following format: 

\begin{quote}
    Two metal balls are the same size but one weighs twice as much as the other. The balls are dropped from the roof of a single story building at the same instant of time. Which of the following describes the time it takes the balls to reach the ground below?

\begin{enumerate}[label=(\alph*)]
    
    \item About half as long for the heavier ball as for the lighter one
    \item About half as long for the lighter ball as for the heavier one
    \item About the same for both balls
    \item Considerably less for the heavier ball, but not necessarily half as long
    \item Considerably less for the lighter ball, but not necessarily half as long

Please answer with a letter (A, B, C, D, or E) and a brief explanation of your reasoning.
\end{enumerate}

\end{quote}

A prior work, in which bar exam questions were administered to ChatGPT, found various tricks that caused it to perform better with multiple-choice questions~\cite{barexam} (the art of tweaking the input to an LLM to optimize its response in this fashion is called ``prompt engineering.") In particular, they found that that, rather than asking ChatGPT for a single answer, it performed better when it was asked to rank its top three choices (though in actuality only it's top choice was scored). However, other similar work~\cite{lawschool} fails to note an effect of this kind, and our own experiments with this alternative prompt structure similarly showed no improvement in the model's overall performance.

We also experimented with two additional procedures. First, we took advantage of  feature offered by ChatGPT which allows the user to request that it ``regenerate response" after it finishes its output. At a gross level, this feature is similar to asking an algorithm for numerically solving some equation to start again but with a different random initial guess. One expects that the results will generally converge to two similar outputs, but perhaps not arrive at exactly the same point. We used this feature as a rudimentary way to explore the ``stability" of ChatGPT's responses, which might in turn be thought of as a proxy for its ``confidence" in its answers. Our very preliminary results based on this experimentation are discussed in Sec.~\ref{sec:stability}. 

Finally, we tested a very different prompt with a very different objective, which we call the ``NOVICE" prompt. In this prompt each question was posed to ChatGPT in the following format: 

\begin{quote}

Please answer the following question as though you were a novice high-school student who has not studied physics and does not yet understand Newton's laws:
    
A large truck collides head-on with a small compact car. During the collision, which of the following is true: 

\begin{enumerate}[label=(\alph*)]

\item the truck exerts a greater amount of force on the car than the car exerts on the truck. 
\item the car exerts a greater amount of force on the truck than the truck exerts on the car. 
\item neither exerts a force on the other, the car gets smashed simply because it gets in the way of the truck. 
\item the truck exerts a force on the car but the car does not exert a force on the truck. 
\item the truck exerts the same amount of force on the car as the car exerts on the truck. 

Just give us your best guess. We know you may not know the correct answer, but we'd like to know which answer makes the most sense to you without any formal physics training. 

\end{enumerate}
\end{quote}
 
We landed on this particular prompt after some trial-and error--see Sec~\ref{sec:novice}. Using this framing, we investigate whether ChatGPT could potentially be useful as a tool for instructors to be able to preview and probe the thinking of a sample ``novice" student while they prepare their teaching materials. 

Initially, we administered the FCI with both the ``BASIC" and ``NOVIVCE" prompts, including stability analysis, To ChatGPT3.5 during the weeks of Feb 13 and Feb 20, 2023. Notably, between our initial experimentation in January and the final administrations of the FCI which generated the results below, there was a significant update to the model which focused on improving its mathematical capabilities, following a series of relatively high-profile examples where users were able to get ChatGPT to espouse manifestly untrue statements about elementary mathematics. Since none of the items in the FCI involve calculations, we think it is unlikely that the update had much impact on our main results. After the release of ChatGPT4, we repeated the same administrations to the new model during the week of March 13th. Results from both models are presented in the remainder of this paper. 

\section{Results}\label{sec:results}

We begin with an analysis of ChatGPT's responses to the 23 usable FCI questions in the modified form described above. For the first, ``BASIC" administration of these problems, we analyze the responses on both a quantitative level, focused purely on its multiple-choice response, and on a qualitative one, by analyzing its stated reason for the answer it chose. 

\subsection{``BASIC" responses}
\subsubsection{Answer Choices}

When prompted with the 23 usable FCI questions using the ``BASIC" prompt format, ChatGPT3.5 gave a correct answer for fifteen of them. It's correct answers are roughly evenly distributed among the unaltered problems and the modified problems which originally contained figures, though it struggles with a certain subset of the modified problems which involve ``spatial reasoning," discussed in more detail in Sec~\ref{sec:fr}.

It is not obvious how we should interpret this number in order to make comparisons to human students taking the FCI. It is tempting to assign ChatGPT a ``score" of $15/23 \approx 65\%$, but it is not entirely clear that this is fair. An argument could instead be made that its ``score" is $15/30 = 50\%$, because of course one important aspect of ``understanding" a physics problem is the ability to synthesize the data being presented across multiple representations~\cite{ThinkingLike}. Since ChatGPT simply cannot comprehend a question that requires reference to a figure, it could be said that it manifestly displays no understanding of that particular item. 

In either case, however, ChatGPT3.5's quantitative performance compares quite adequately with the post-test results that are typical for students taking the FCI at the end of their first semester of college-level physics. We can make a direct comparison to one of the author's previous classes in 2018, in which 415 students took the FCI at the end of the term, and produced the distribution of scores found in figure~\ref{fig:posthist}. In that distribution, the median student score was a $56\%$, meaning that depending on how charitably one treats its partial results, ChatGPT was either just below the performance of a typical student or else perhaps a nontrivial cut above average. As an alternative perspective, the median letter grades of the students scoring around a 50\% or a 65\% were a B- and a B+, respectively.

\begin{figure}[h]
    \centering
    \includegraphics[width=0.4\textwidth]{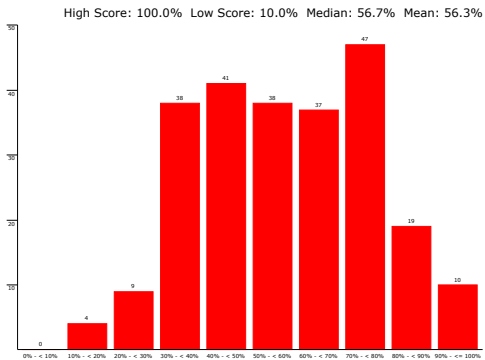}
    \caption{ The distribution of FCI scores at the end of a semester of college physics at a large public R1 university. ChatGPT3.5's performance using ``BASIC" prompting would put it either in the bin containing the median (50\%) or in the bin just above (65\%) }
    \label{fig:posthist}
\end{figure}

When we repeated the analysis using ChatGPT 4, we found that the model now responded ``correctly" to 22 of the 23 questions. Even more strikingly, the question which it ``missed" was item 26, which asks about the behavior of a box being pushed against friction. The answer which it chooses, "D," would in fact be correct if air resistance were to be considered non-negligible, and the problem text does not specify this condition either way. Students are presumably expected to make the standard (and physically reasonable assumption) that the air resistance on a box being pushed by a human is a small effect compared to other forces in the problem. ChatGPT perhaps betrays some ``inhuman" behavior by not making a similar assumption. Nevertheless, when we prompted the model in a new instance with the same question and added the text ``assume air resistance is negligible," it gives the correct answer here as well. 

Needless to say, the performance of ChatGPT4 compares to the very uppermost decile of performance by physics students in our post-test data, and of course its near-perfect responses no longer show and weakness for problems that originally relied on figured and/or spatial reasoning.

\subsubsection{Free Responses}\label{sec:fr}

As part of the ``BASIC" prompt, we asked ChatGPT to provide ``a brief explanation of [its] reasoning." This gives us an additional dimension along which to assess its performance, albeit a qualitative one. 

As a baseline, we read through each response and made a decision about whether the answer was completely accurate, which is the level of performance we would expect from an expert physicist considering such introductory concepts. In this determination we disregarded tone and focused only on content. So, ChatGPT was not penalized if its correct explanation sounded stilted, in a way that suggested machine learning rather than a human intelligence behind the response. On the other hand, we marked ``wrong" any response that contained an error which a trained physicist would not reasonably have made, no mater how ancillary the error was to ChatGPT's analysis. Certainly, these determinations were simply a judgement call on the part of the author and of course were not blind to the fact that the responses were generated by a language model. However, we feel that in most cases the determinations were relatively black-and-white, since they centered on a decision about whether the statements made in the explanation were factually true. 

In our judgement, 10 of ChatGPT3.5's 23 free responses ($43\%$) fully passed this ``expert test." These responses were typically relatively terse, invoking the correct constellation of concepts, and often making explicit note of facts or concepts which ruled out wrong answers as well as the facts which supported the correct one. For example, item 13 of the FCI says 

\begin{quote}

``A boy throws a steel ball straight up. Consider the motion of the ball only after it has left the boy's hand but before it touches the ground, and assume that forces exerted by the air are negligible. For these conditions, what are the force(s) acting on the ball? 
    \begin{enumerate}[label=(\alph*)]

    \item A downward force of gravity along with a steadily decreasing upward force. 
    \item A steadily decreasing upward force from the moment it leaves the boy’s hand until it reaches its highest point; on the way down there is a steadily increasing downward force of gravity as the object gets closer to the earth. 
    \item An almost constant downward force of gravity along with an upward force that steadily decreases until the ball reaches its highest point; on the way down there is only a constant downward force of gravity. 
    \item An almost constant downward force of gravity only. 
    \item None of the above. The ball falls back to ground because of its natural tendency to rest on the surface of the earth. 
    \end{enumerate}

\end{quote}

Note the absence of terms like ``projectile motion," which might cause a language model (or a student!) to mechanically regurgitate the fact that only the force of gravity is acting. Despite this challenge, ChatGPT'3.5s response, which was representative of its performance on these ten ``entirely correct" free responses, was as follows: 

\begin{quote}

(D) An almost constant downward force of gravity only.

After the steel ball leaves the boy's hand, the only force acting on it is the force of gravity, which is directed downwards. This force is almost constant throughout the motion of the ball, with only slight variations due to changes in the distance between the ball and the center of the Earth.
There is no upward force acting on the ball, as the ball is not being propelled upwards once it leaves the boy's hand. As a result, the ball will continue to rise until the force of gravity slows it down and eventually reverses its direction, causing it to fall back to the ground.
Therefore, the correct statement is (D) An almost constant downward force of gravity only.

\end{quote}

In addition to identifying the correct force, it makes specific note of the absence of any ongoing propulsive force (For comparison, in the Fall '18 post-test student data, 38\% of students chose one of the responses which includes an upward force during the motion). We note also that ChatGPT3.5 goes out of its way to comment on and correctly explain why the correct answer choice has the qualifier ``almost-constant" on the force of gravity. This level of commentary is in fact deeper than we would typically expect of a first-semester physics student, at least unprompted. In fact, the inclusion of this detail in the question strikes us as the kind of thing an instructor does to make sure that their questions are airtight so that the strongest students have no room to complain or overthink. 

Of course, not all of ChatGPT3.5's responses reach this level of accuracy. There were five free-responses which we judged to be more correct than not (in fact, typically almost entirely correct), but containing at least one patently untrue statement which simply could not have been made by an expert physicist. These minor errors were often ``surprising" in the sense that they did not necessarily represent the kinds of classic misapprehensions we have come to expect from new students on similar questions. And the errors were typically ``unforced" in the sense that they did not appear in the portion of the explanation that was strictly necessary to identify the correct answer. For example, consider item 22 of the FCI, which asks about a rocket in deep space that was moving horizontally from left to right at a constant speed until its engine produced a period of constant upward force. The question asks about the speed of the rocket while the engine is on (whether it is increasing, decreasing, constant, etc). 

ChatGPT3.5 correctly identifies that the speed is increasing while the engine is on, but says the following: 

\begin{quote}
    Since the rocket is subject to no outside forces, its motion is determined solely by the forces it experiences from its engine. When the engine is turned on, the rocket is subject to a constant thrust that is perpendicular to its initial motion. This thrust produces an acceleration that is also perpendicular to the rocket's initial motion. As a result, the rocket moves along a circular path, with its speed increasing as it moves from point "b" to point "c". Therefore, the answer is (B) its speed is continuously increasing.
\end{quote}

All of the elements of the correct chain of reasoning are present: (1) the net force in this case is equal to the force of the engine (2) when the engine is on, the net force is nonzero (3) a nonzero net force causes an acceleration and (4) the acceleration is not directed against the motion and hence it will cause the velocity to increase. But ChatGPT3.5 make an unforced error by trying to describe the nature of the rocket's motion during this time, even though this was not part of the question. It identifies the rocket's motion as ``circular," perhaps because in its corpus of physics there is a strong association between discussions of circular motion and ``forces which are perpendicular to the direction of motion." Missing the obvious detail that this particular force is only \textit{initially} perpendicular to the motion, it draws an incorrect conclusion. And the mistake is as surprising as it is unnecessary: we do not imagine that many freshman physics students are picturing a rocket which becomes locked into something like cyclotron motion when they read this problem, even though many of the found it difficult (57\% chose incorrectly on the Fall '18 post-test).

These five responses correspond with the other five problems for which ChatGPT3.5 chose the correct multiple-choice answer, although we graded these responses separately and it was not a foregone conclusion that it would coincide in this way. 

After 10 fully-correct responses and five nearly-correct responses that nevertheless contain unambiguous errors, ChatGPT3.5 gives eight responses which are nowhere near the mark. These responses typically contain an assortment of words and phrases which are genuinely relevant to answering the problem. Often, entire sentences form accurate statements about the underlying physics. But the ideas are linked up in ways which are grammatically correct but often logically unreasonable or inconsistent, and the responses typically also contain discussions of details that are wholly irrelevant, intermixed on equal footing with the key concepts. The analyses may swerve unpredictably onto different tracks, or arrive at conclusions entirely at odds with the preceding arguments, as if the model had no idea what it had just said, and yet it continues confidently to claim an inescapable conclusion. They certainly do \textit{not} project anything like understanding--quite to the contrary, if given by a human they would be taken as specific evidence that they are regurgitating memorized words and do not know what they are talking about. 

In defense of ChatGPT3.5, half of the eight problems in which the free-response is wide of the mark are problems that originally contained figures. In particular, these four problems are the ones in which the answers rely most on spatial reasoning and on establishing some sense of direction, and ChatGPT3.5 seems simply unable to recognize the directional system being used.


Consider for example its response to item seven: the problem regarding a ball released during circular motion which was described above. In our wording of the question, the ball is moving north at the time its string is cut and it is allowed to move off tangent to the circle, so the correct answer about its subsequent motion should be that it travels in a straight line to the north. ChatGPT3.5 comes close to understanding this, but appears confused about two points relating to the directional system, explaining:  

\begin{quote}
     When the string breaks, the ball will continue to move in a straight line tangential to the circular path it was following at that moment. Since the ball was moving north, the tangential velocity of the ball will be directed to the east.

    However, the force of gravity will also be acting on the ball, pulling it downward towards the Earth. As a result, the ball will follow a curved path towards the west as it falls. 

    Therefore, the correct statement is (A) It will initially travel north, but will quickly begin to curve to the west. 
\end{quote}

Aspects of this response are perfectlly reasonable. The first sentence is entirely true, and so is the third-- though most students answering this problem would ignore this vertical motion and answer only with respect to its horizontal motion, as the figure suggests. But from these two true statements, ChatGPT3.5 is unable to draw any correct conclusions about the real world direction, making the bizarre leaps from ``tangential" to ``east" and ``downward" to ``west." One possible reason for this is that while there are good reasons to expect terms like ``tangential" and "gravity" to be preferentially associated in the training corpus with discussion of balls released from circular motion, there is no reason that a particular direction like "north" would be similarly associated. Hence, in trying to model how a human would talk about this problem, ChatGPT3.5 gravitates correctly to details that all expert humans would mention, but then has to pick randomly to fill in a direction, because presumably all the cardinal directions are represented roughlly equally in the training text. ChatGPT3.5 has succeeded at saying something that conveys general understanding based on general patterns, but failed to show specific understanding of the details in the instance at hand.

ChatGPT4, by contrast, gave no answers in this ``highly confused" category. We assessed 20 of its 23 free responses as being ``exactly correct," as one would expect from an expert physicist. The GPT4 model is notably more concise in general than its predecessor, and this brevity may have helped it avoid some of the ``unforced errors" seen previously. But it also succeeded on nearly all of the problems for which ChatGPT3.5 was completely confused. Consider item seven, for example: the ball released from circular motion which ChatGPT3.5 assessed correctly in general terms but failed to answer correctly because it could not identify which direction was ``tangential" in the particuar case at hand. ChatGPT4 instead gives a correct answer with perfectly correct reasoning: 

\begin{quote}
    When the string breaks, the ball will continue in the direction it was moving at the exact moment the string broke, due to its inertia. Since the ball was moving north when the string broke, it will continue to travel north in a straight line.
\end{quote}

Like ChatGPT3.5, ChatGPT4 has identified a correct general concept to invoke (inertia), but has also succeeded at drawing a specific inference in this case (note that this answer was ``stable" and not a lucky guess; see below). Without more public information about the improvements in ChatGPT4 and the workings of its head layers in this case, we cannot know how it arrives here. But it seems plausible that it has been an able to extract a deeper level of patterns from its traning text. The base pattern is that ``inertia" and ``motion in the tangential direction" commonly discussed in this context. But with in that pattern is a deeper one: in examples where humans discuss such questions, the solution typically refers to the exact same direction-- left, north, +y etc-- which was referenced in the problem next to words referring to its motion at the motion the ball is released. Whether or not this is exactly why ChatGPT4 performs better on this problem, it is a good example of how increasing the models inferential depth can allow it to project an apparently deeper understanding physics without having been``taught" any more explicit physics concepts.

ChatGPT4's three remaining response were all ``nearly perfect," with solid reasoning and incidental minor errors. For example, in response to item 16, which asks about the forces on an elevator moving at constant velocity, it says that there must be no net force on the elevator and hence that ``any forces acting on the object must come as equal and opposite pairs." This leads it to a correct conclusion that in this particular case that the upward tension from the elevator cable has the same magnitude as the downward weight force. But an expert analyzing the problem would not quite have made the statement that ``any forces" acting MUST come as pairs-- an elevator with two cables, for example, could still move at constant velocity without any two forces existing as an equal and opposite set. But this level of ``error," which represents ChatGPT4's worst performance, is far more minor than the worst mistakes of ChatGPPT3.5.

\subsubsection{Stability}\label{sec:stability}

Because there is a probabilistic element inherent in ChatGPT's operation, we make a brief study of the ``stability" of the models' responses by experimenting with two types of perturbations. This investigation is suggestive but very preliminary; deeper work could certainly be done on this topic.

First, we make use of ChatGPT's ``regenerate response" button, which allows us to simply request that the model start over its process with the exact same prompt, albeit with some variation in whatever internal parameters it uses to traverse its vast network of possible responses. This is analogous to asking a numerical equation solver to start over with a different random initial guess, and just as in that case, one would hope that a stable method would consistently produce comparable outputs.

For this first, initial probe of the Chatbot's stability, we regenerated each response just three times, but already some patterns seemed clear. The only cases where ChatGPT3.5 consistently changed its answers between regenerations were a subset of the eight questions for which we judged that its written responses were completely confused. There were two other cases of correct answers where the model occasionally changed its mind when the responses were regenerated, but in those cases it stuck with the correct answers a clear majority of the time. This pattern is perhaps somehow comforting or impressive for the question of ChatGPT's ability to project understanding: when it knows the answer, it knows the answer with some stability. When it is flying by the seat of its pants, it also does not care what destination it flies to. 

The other form of perturbation which should be explored is perturbations to the input, rather than to the initial starting conditions of the model themselves. Of course, because we cannot control the inner workings of the model and put it back in the exact same internal state every time, it is not possible to disentangle these two. Nevertheless, we attempted for a subset of the problems to feed ChatGPT3.5 variations on the same questions but with irrelevant words and sentence structures switched around (e.g. ``A boy throws a steel ball straight up" becomes ``a rock is tossed straight upward by a girl). Once again, we found that this generally did not affect the response if the model got the initial problem right, but it could change the responses for the problems it was already getting completely wrong.

Our discussion here has focused on ChatGPT3.5 because we found that ChatGPT4's responses were completely insensitive to either of the perturbation types described above. Perhaps this is not surprising: ChatGPT3.5's responses were only unstable on problems where its responses were confused and incoherent, and ChatGPT4 never reached that level of confusion.

\subsection{"NOVICE" Responses}\label{sec:novice}

This paper is focused on the question of whether ChatGPTs behavior is consistent with ``understanding" introductory physics, but part of our motivation for asking that question is to understand how it may affect physics classrooms and physics pedagogy more broadly. One way it might be of use to an instructor would be as a way to gain insight into the unfamiliar mind of a novice physics student. If ChatGPT were able to successfully answer questions in a manner that plausibly mimicked typical student mental models~\cite{MentalModels} it might be of great value in testing and preparing lesson plans, refining exam questions, etc. 

The question of whether ChatGPT can do this is the worthy subject of its own project, but we make an initial stab at it here to establish some baseline results and investigate whether such deeper work is likely to bear fruit. To do this, we gave the modified FCI questions to ChatGPT again, but this time using a prompt which asks it to answer as though it had not yet studied any introductory physics. This prompt was developed through trial and error using a small subset of FCI problems for which we had strong evidence from our own students' prior pretests about how they might answer. These were items 4 and 26 from the FCI. Item 4 tests Newton's 3rd law in the unintuitive context of an asymmetric collision between a truck and a car. Prior to studying physics~\footnote{and, sadly, in a good portion of the time \textit{after} studying physics}, many students (nearly 75\% in our fall '18 pretest) believe that the truck will exert a greater force on the car than vice versa. We found that initially, even when we prompted ChatGPT to answer as though it had not studied Newton's Laws, it continued to give the expert answer that the forces were equal. It was only when we modified the prompt to include a reminder of our desired reference frame both before and after the question text itself that we saw it give the infamous novice answer described above. We saw similar behavior in question 26, which tests concepts about the balance of forces in the context of kinetic friction. 

The results from ChatGPT3.5 were mixed. To quantitatively evaluate the performance, we scored its answers on a ``key" comprising all of the most common responses given by students in our Fall '18 pretest data, 10 of which were already correct answers. ChatGPT's answers to questions with the ``NOVICE" prompt matched these most-common student answers on 11 occasions out of 23.  In particular, seven of these 11 matches came in situations where the most common pretest answer was itself a correct answer. While it is not inherently a bad to match in these cases, because an educator using ChatGPT to test out possible student responses would want to know about cases where even an untrained student is likely to get the question right, it is discouraging to see comparatively few cases where the plurality student pretest opinion was wrong and ChatGPT was also able to identify the distractor they would find most compelling. 

Curiously, the problem was not that ChatGPT did ``too well" on the problem set. In fact, ChatGPT's ``NOVICE" responses to the 23 modified FCI questions were right only nine times. There were only three such occasions where ChatGPT got a question right which the pretest plurality got wrong. But clearly, in the remaining cases, it did not agree with the plurality about which wrong answer seemed most intuitive. 

This binary, agree-or-disagree framework is a little unfair to ChatGPT. Imagine a scenario where 50\% of students choose wrong answer ``A," 49\% choose wrong answer ``B," and only 1\% choose the correct answer ``C." If ChatGPT were to offer answer ``B" when roleplaying as a student, it would hardly represent a glaring failure. To capture some of this nuance, we propose a simple measure where ChatGPT is given a point for each answer, weighed by the fraction of students from our Fall '18 pretest who chose that answer (so, half a point if 50\% of students chose the same answer). We exclude problems which the plurality of students got right, since in such cases we cannot tell anything about whether ChatGPT is understanding novice thinking or simply agreeing with their conclusion as an expert would. Summing these scores and then normalizing by the equivalent score of the student key itself (which represents the maximum possible number of points ChatGPT could score) gives ChatGPT3.5 a score of 63. Note that if we run this calculation for the average performance of someone guessing randomly, for this particular distribution of student responses, the minimum score would be 58. Hence, ChatGPT3.5's performance is only a little better than random here. 

To our surprise, the performance of the otherwise-superior ChatGPT4 is substantially \textit{worse} on this task. In fact, the GPT-4 model proved completely incapable of straying from the ``correct" answers in response to this new prompt, and a variety of others which were tested. Its score using the same metric above was a 36-- i.e., a person randomly trying to guess what a novice would choose would outperform it, because the model insisted on choosing correct answers, and these were rarely chosen by novice students in practice. 

Examining the free-responses of both ChatGPT3.5 and ChatGPT4 under the ``Novice" prompt gives some insight into the reason it wasn't better able to mimic the behavior of an untrained student. For ChatGPT3.5, a significant majority of its explanations referenced concepts like forces and accelerations that would likely not be the basis for a true novice's analysis of the problem. In fact, it routinely mentioned specific cases Newton's laws by name, despite the explicit instruction that it should answer as a student who ``does not yet know or understand" them. It seemed instead to be answering from the perspective of a student who knew about Newton's laws, but who was having trouble applying them correctly. ChatGPT4's responses differed just slightly: it was better about avoiding explicit mention of Newton's laws, but would simply claim that its intuition coincided with the correct Newtonian framework using nontechnical terms. For example, analyzing an object moving at a constant velocity, it would say, ``the upward pull must be just as big as the downward pull, because it makes sense that the forces must be balanced in a case like this."

The picture that emerges  from both models is that of a traditional professor, who understands the physics well but struggles to recognize their students' difficulties because they can scarcely remember what it was like not to know the concepts by heart. We leave it for future work to see whether the ``novice" performance could improve if we further refine the prompt, or perhaps engage in more back-and-forth with the model. However, we conjecture that it may be in the very nature of a LLM trained on a broad corpus of human text that it struggles to answer questions with a self-imposed blind spot. After all, cases of experts discussing physics correctly presumably represent the vast majority of cases in which physics was discussed in ChatGPT's training text. And there were likely some instances in which students mentioned Newton's laws, identified that they did not know how to use them, and proceeded to use them incorrectly, which is closer to what we saw in the responses from ChatGPT3.5. But seems much less likely to have seen cases where students mentioned Newton's laws by name, but only for the purpose of saying that they had never heard of Newton's laws and to describe their reasoning in the absence of them. As such, ChatGPT might inevitably struggle to identify cases where students were reasoning with a pre-Newtonian mindset based on an instruction like ``you do not know or understand Newton's laws" absent a training set specifically designed for this purpose. 


\section{Conclusions}\label{sec:conclusions}

Our paper asks whether ChatGPT ``can project an understanding of introductory physics," meaning, ``can it display behaviors consistent with having an underlying understanding of kinematics and Newtonian dynamics, whether or not such underlying understanding actually exists?" For Chat GPT3.5, the answer appears to be ``yes locally, not globally." In other words: in some isolated cases it responded to items from the FCI designed to test for conceptual mastery of introductory physics exactly the way an expert physicist might, despite (as far as we can tell) having never seen the question before and not having any specific programming dedicated to ``doing physics."  ChatGPT3.5 displays this ability about the same percentage of the time as a B- or B-level student taking a college physics course. This is considerably more proficiency than many physicists would have predicted was imminent just a few years ago. On the other hand: when the GPT3.5 mask slips, what it reveals is so clearly devoid of understanding that it spoils the charade. If a student submitted work which showed perfect mastery in one place and complete incoherence on the same set of topics immediately after, we would suspect cheating. ChatGPT may not be ``cheating" per se, but it is not on the whole performing in a manner that one projects full-fledged expertise. 

The story of ChatGPT4 is quite different. There are many aspects of true ``understanding" -- such as the ability to engage in metaphor~\cite{metaphor} or make use of multiple representations~\cite{representations}--which differentiate experts and novices and which we do not attempt to test here. As such, perhaps we still cannot claim that ChatGPT4 ``globally" projects understanding of introductory physics. But for whatever subset of conceptual understanding is captured by the FCI (a tool built precisely to capture some aspect of deeper comprehension) ChatGPT4 is completely capable of responding like an expert, making only one mistake in its responses (and even then, making a mistake that a nitpicking expert might be inclined to defend).

There are at least four valuable questions to be asked in light of this performance, and we provide speculative discussion here in anticipation of further work. First, one could ask what ChatGPT's performance can tell us about the nature of some of the standardized assessment tools used in PER. One answer appears to be that it is important to treat these instruments as a whole, and not place too much emphasis on any single item. Assessments like the FCI were designed intentionally to probe central concepts along various different dimensions and with slight variations in the presentations of the concepts. ChatGPT3.5's oscillation between expertise and vacuousness underscores the necessity of this. The results also remind us that there is yet another distinction to be drawn between ``understanding" a concept, in the sense of ``being able to apply it like an expert," and \textit{believing} a concept, in the sense of having an internal satisfaction that the concept is true because it should necessarily be true. This distinction is significant also among our students: compare how relatively easy it is for a student to learn that ``both forces in the pair are always equal" in the context of Newton's Third law with how few students actually feel that this fact is reasonable and intuitive when they first encounter it~\cite{TeachingN3}.

Second, given its level of apparent understanding, one could investigate ChatGPT's potential use by students, because it (or tools like it) will soon be used by students in our classroom, which is already happening in other fields of study~\cite{GPTHomework}. At the moment, a student relying on ChatGPT3.5 might be easy enough to spot, given its inconsistency. But because its faulty responses stand out so much, an enterprising student could conceivably ask it for help with every question and then learn the signs of when to disregard its answers. And of course, while ChatGPT4 is currently available only on a subscription basis and with substantial rate limitations that will prevent it from being used broadly by every student in an intro physics class, these restrictions are surely likely to recede with time. All this is to say, even at its current ability level, ChatGPT3.5 threatens the integrity of things like take-home tests, exam corrections, lab reports, written homework assignments, and online problem sets, at least in their current form. ChatGPT4's performance is significantly stronger still.

Third, we considered the possible role that ChatGPT could have as an in-class teaching tool. In some fields where its responses have a higher success rate, like computer science, some faculty are encouraging their students to come to class with a ChatGPT tab open, and to use it to ask brief clarifying questions, or find the bugs in their sample code, so that many more of these minor questions can be handled than a professor could hope to field on their own~\cite{ClassWithChatGPT}. At this juncture, we clearly do not recommend using ChatGPT3.5 this way in a physics classroom, as it's hit rate for correct answers and correct explanations is simply too low, and the danger of confusing a student (or undermining their confidence) by giving them information that is only statistically trustworthy is too great. However, ChatGPT4 seems likely to have crossed the crucial threhsold here, and as it becomes widely available it may indeed be possible for a student to use it as a ``copilot" to aid their understanding in the same way that students currently might use tools like calculators and computer algebra systems to help them track the mathematics in their physics courses. We do not opine at this juncture on whether this kind of use would be desirable, except to say that the strength and ubiquity of ChatGPT and its successors might make it inevitable, in which case it may be preferable for educators now to be identifying the best ways to do so. 

Finally, we asked about using ChatGPT as a tool to support physics instructors \textit{outside} the classroom, by assisting with preparation. Here, it is clear that ChatGPT is \textit{not} able to play at least one of the roles we had initially conjectured: it cannot conjure a useful simulacrum of a novice mindset in a way that would allow an educator to test out their various examples and analogies to see how they might land with their students. But there are still some other ways it could be useful. It's partial understanding and relatively literal mode of interpretation, for example, make it an attractive tool to playtest possible exam questions to root out glitches, loopholes, or overdetermined facts. ChatGPT might also function as a tool to help with the challenging process~\cite{examQs}  of drafting of new exam or homework questions in the first place. 

All of these motivating questions immediately suggest areas for follow-up projects. We intend to pursue some ourselves, and encourage others to do so as well. And we must note that, while history contains faulty predictions about the timeline of AI development in both directions~\cite{AIPredictions}, the current pace and nature of the field suggests it will continue to advance rapidly, as evidenced by the dramatic leap in performance from ChatGPT3.5 to ChatGPT4 which came about even within the time it has taken to prepare this work for publication. Indeed, the self-reinforcing nature of the field as AI models learn to train themselves, coupled with the current exponential growth of computing power, suggests that the field as a whole could advance at an exponential or even superexponential rate~\cite{Superexponential}. Concepts which at the time of this writing can be dismissed as being beyond even ChatGPT4's capabilities may be old news by the time the reader finds this. This is all the more reason why continued work in this area is urgently needed. 

\section*{Acknowledgement}
The author thanks Noah Finkelstein for useful discussions regarding the scope and framing of this work, and Mark Kissler, M.D.,  of the University of Colorado Hospital for suggesting the phrase ``project understanding" as a way to clarify the central research question explored here. 

\bibliographystyle{unsrt}
\bibliography{references}

\begin{thebibliography}{10}

\bibitem{Gerd}
Gerd Kortemeyer.
\newblock Could an artificial-intelligence agent pass an introductory physics
  course?
\newblock {\em arXiv preprint arXiv:2301.12127}, 2023.

\bibitem{OpenAI}
OpenAI.
\newblock Chatgpt: Optimizing language models for dialogue, 2022.

\bibitem{TransformerOriginal}
Ashish Vaswani, Noam Shazeer, Niki Parmar, Jakob Uszkoreit, Llion Jones,
  Aidan~N Gomez, {\L}ukasz Kaiser, and Illia Polosukhin.
\newblock Attention is all you need.
\newblock {\em Advances in neural information processing systems}, 30, 2017.

\bibitem{Pretraining}
Long Ouyang, Jeff Wu, Xu~Jiang, Diogo Almeida, Carroll~L Wainwright, Pamela
  Mishkin, Chong Zhang, Sandhini Agarwal, Katarina Slama, Alex Ray, et~al.
\newblock Training language models to follow instructions with human feedback.
\newblock {\em arXiv preprint arXiv:2203.02155}, 2022.

\bibitem{FewShot}
Tom Brown, Benjamin Mann, Nick Ryder, Melanie Subbiah, Jared~D Kaplan, Prafulla
  Dhariwal, Arvind Neelakantan, Pranav Shyam, Girish Sastry, Amanda Askell,
  et~al.
\newblock Language models are few-shot learners.
\newblock {\em Advances in neural information processing systems},
  33:1877--1901, 2020.

\bibitem{Importance1}
Eva~AM van Dis, Johan Bollen, Willem Zuidema, Robert van Rooij, and Claudi~L
  Bockting.
\newblock Chatgpt: five priorities for research.
\newblock {\em Nature}, 614(7947):224--226, 2023.

\bibitem{Importance2}
The PyCoach.
\newblock Chatgpt: The end of programming (as we know it), Dec 2022.

\bibitem{GPT4}
OpenAI.
\newblock {GPT}-4 technical report.
\newblock {\em arXiv preprint arXiv:2303.08774}, 2023.

\bibitem{wharton}
Christian Terwiesch.
\newblock Would chat gpt3 get a wharton mba? a prediction based on its
  performance in the operations management course.
\newblock {\em Mack Institute for Innovation Management at the Wharton School,
  University of Pennsylvania. Retrieved from: https://mackinstitute. wharton.
  upenn. edu/wpcontent/uploads/2023/01/Christian-Terwiesch-Chat-GTP-1.24. pdf
  [Date accessed: February 6th, 2023]}, 2023.

\bibitem{lawschool}
Jonathan~H Choi, Kristin~E Hickman, Amy Monahan, and Daniel Schwarcz.
\newblock Chatgpt goes to law school.
\newblock {\em Available at SSRN}, 2023.

\bibitem{barexam}
Michael Bommarito~II and Daniel~Martin Katz.
\newblock Gpt takes the bar exam.
\newblock {\em arXiv preprint arXiv:2212.14402}, 2022.

\bibitem{BarExam4}
Daniel~Martin Katz, Michael~James Bommarito, Shang Gao, and Pablo Arredondo.
\newblock Gpt-4 passes the bar exam.
\newblock {\em Available at SSRN 4389233}, 2023.

\bibitem{FCI}
David Hestenes, Malcolm Wells, and Gregg Swackhamer.
\newblock Force concept inventory.
\newblock {\em The physics teacher}, 30(3):141--158, 1992.

\bibitem{commoncrawl}
Common crawl.
\newblock Faq.

\bibitem{Note1}
There may be an exception for raw mathematics--see Methods section.

\bibitem{6facets}
Grant~P Wiggins and Jay McTighe.
\newblock {\em Understanding by design}.
\newblock Ascd, 2005.

\bibitem{minsky1982}
Marvin~L Minsky.
\newblock Why people think computers can't.
\newblock {\em AI magazine}, 3(4):3--3, 1982.

\bibitem{Noah}
Julian~D Gifford and Noah~D Finkelstein.
\newblock Categorical framework for mathematical sense making in physics.
\newblock {\em Physical Review Physics Education Research}, 16(2):020121, 2020.

\bibitem{CarlKathy}
Carl Wieman and Katherine Perkins.
\newblock Transforming physics education.
\newblock {\em Physics today}, 58(11):36, 2005.

\bibitem{elby}
Andrew Elby.
\newblock Helping physics students learn how to learn.
\newblock {\em American Journal of Physics}, 69(S1):S54--S64, 2001.

\bibitem{patternmatching}
Jonathan Tuminaro and Edward~F Redish.
\newblock Understanding students’ poor performance on mathematical problem
  solving in physics.
\newblock In {\em AIP Conference Proceedings}, volume 720, pages 113--116.
  American Institute of Physics, 2004.

\bibitem{WilcoxPollock}
Bethany~R Wilcox and Steven~J Pollock.
\newblock Investigating students’ behavior and performance in online
  conceptual assessment.
\newblock {\em Physical Review Physics Education Research}, 15(2):020145, 2019.

\bibitem{RLHF}
Paul~F Christiano, Jan Leike, Tom Brown, Miljan Martic, Shane Legg, and Dario
  Amodei.
\newblock Deep reinforcement learning from human preferences.
\newblock {\em Advances in neural information processing systems}, 30, 2017.

\bibitem{blogpost}
Alexa Steinbrueck.
\newblock Can chatgpt do image recognition?, Jan 2023.

\bibitem{HalfFCI}
Jing Han, Lei Bao, Li~Chen, Tianfang Cai, Yuan Pi, Shaona Zhou, Yan Tu, and
  Kathleen Koenig.
\newblock Dividing the force concept inventory into two equivalent half-length
  tests.
\newblock {\em Physical Review Special Topics-Physics Education Research},
  11(1):010112, 2015.

\bibitem{ThinkingLike}
Alan Van~Heuvelen.
\newblock Learning to think like a physicist: A review of research-based
  instructional strategies.
\newblock {\em American Journal of physics}, 59(10):891--897, 1991.

\bibitem{MentalModels}
Donald~A Norman.
\newblock Some observations on mental models.
\newblock {\em Mental models}, 7(112):7--14, 1983.

\bibitem{Note2}
and, sadly, in a good portion of the time \protect \textit {after} studying
  physics.

\bibitem{metaphor}
Fredrik Jeppsson, Jesper Haglund, and Tamer~G Amin.
\newblock Varying use of conceptual metaphors across levels of expertise in
  thermodynamics.
\newblock {\em International Journal of Science Education}, 37(5-6):780--805,
  2015.

\bibitem{representations}
Patrick~B Kohl and Noah~D Finkelstein.
\newblock Patterns of multiple representation use by experts and novices during
  physics problem solving.
\newblock {\em Physical review special topics-Physics education research},
  4(1):010111, 2008.

\bibitem{TeachingN3}
CH~Poon.
\newblock Teaching newton's third law of motion in the presence of student
  preconception.
\newblock {\em Physics Education}, 41(3):223, 2006.

\bibitem{GPTHomework}
J{\"u}rgen Rudolph, Samson Tan, and Shannon Tan.
\newblock Chatgpt: Bullshit spewer or the end of traditional assessments in
  higher education?
\newblock {\em Journal of Applied Learning and Teaching}, 6(1), 2023.

\bibitem{ClassWithChatGPT}
Thomas Rid.
\newblock Five days in class with chatgpt, Jan 2023.

\bibitem{examQs}
James~H Smith and Alfred~G Costantine.
\newblock Writing better physics exams.
\newblock {\em The Physics Teacher}, 26(3):138--144, 1988.

\bibitem{AIPredictions}
Stuart Armstrong, Kaj Sotala, and Se{\'a}n~S {\'O}~h{\'E}igeartaigh.
\newblock The errors, insights and lessons of famous {A}{I} predictions--and
  what they mean for the future.
\newblock {\em Journal of Experimental \& Theoretical Artificial Intelligence},
  26(3):317--342, 2014.

\bibitem{Superexponential}
Nick Bostrom.
\newblock {\em Superintelligence: Paths, Dangers, Strategies}.
\newblock Oxford University Press, Oxford, 2014.

\end{thebibliography}
\end{document}